  \documentstyle[12pt, aaspp]{article} 
  \def\kms{km s$^{-1}$} 
   
  \def\and{$\&$ } 
  \def\rod{Rodr\'\i guez} 
   
  \def\asec{$^{\prime\prime}$} 

  \received{1 July 2003} 

\begin{document} 

  \title{Broad Recombination Line Objects in W49N on 600 AU Scales}

  \author{C. G. De Pree\altaffilmark{1}, D. J. Wilner\altaffilmark{2}, A. J. Mercer\altaffilmark{1}, L. E. Davis\altaffilmark{1},  W.M. Goss\altaffilmark{3}, S. Kurtz\altaffilmark{4}}.


  \altaffiltext{1}{Department of Physics and Astronomy, Agnes Scott College, 141 E. College Ave., Decatur, GA
  30030} 
  \altaffiltext{2}{Center for Astrophysics, 60 Garden Street, Cambridge, MA 02138} 
  \altaffiltext{3}{National Radio Astronomy Observatory, P.O. Box 0, Socorro, NM 
  87801} 
  \altaffiltext{4}{Centro de Radioastronom\'\i a y Astrof\'\i sica, UNAM, Morelia, Mexico} 
    
  \begin{abstract} 
High resolution 7 mm observations of the W49N massive star forming region 
have detected recombination line emission from the individual
ultracompact (UC) HII regions on 50 milliarcsecond 
(600 AU) scales. These line observations,
combined with multifrequency,
high-resolution continuum imaging of the region at 7 mm 
(VLA) and at 3 mm and 1 mm (BIMA), indicate that five to seven of the 
eighteen 
ultracompact sources in W49N are broad
recombination line objects (BRLOs) as described by Jaffe \& Martin-Pintado (1999). BRLOs
have both broad radio recombination lines ($\Delta$V$>$60 \kms) 
and rising spectra (S$_{\nu}\sim\nu^{\alpha}$), with $\alpha$ values
greater than 0.4.
The broad line widths of the H52$\alpha$ line
are probably related
to motions in the ionized gas rather than pressure broadening. 
A number of models have been proposed to explain the long lifetime of UC HII regions, including
the photoevaporated 
disk model proposed by Hollenbach et al.
(1994). This model can also explain the broad lines,
rising spectra and bipolar morphologies of some sources. 
We suggest$-$based on line and continuum 
observations as well as source morphology$-$that in a subset
of the W49N ultracompact sources we may be
observing ionized winds that arise from circumstellar disks.\end{abstract} 



  %
  %

  \newpage 
  \section{INTRODUCTION} 
The W49A star-forming region is among the most luminous in the
Galaxy (10$^7$ L$_\odot$) and is located within one of its most massive
giant molecular clouds (10$^6$ M$_\odot$). While W49A is a distant region 
(located
at about 11.4 kpc), the large number, high luminosity and wide 
morphological variety of its HII regions
make it a valuable region to study
high mass star formation, as long as observations have sufficient
angular resolution. The W49 North (W49N) cloud core within
W49A contains over a dozen ultracompact (UC) HII regions
powered by OB-type stars arranged in a 2 pc diameter ring (Dreher
et al. 1984; Dickel \& Goss 1990). The
discovery of so many compact sources within W49N was one of the
origins of the UC HII region "lifetime problem". That is, the observed number
of UC sources seems to exceed the numbers expected based on a
free expansion of the ionized gas (Dreher et al. 1984; Wood \&
Churchwell 1989). A variety of models have been proposed to
account for this discrepancy, and were recently summarized in Jaffe 
\& Martin-Pintado (1999; hereafter JMP).

A recent millimeter survey of UC HII regions (JMP) indicates
that about 30\% of them are characterized by (1) rising spectral indices 
($\alpha>$0.4), and (2) broad radio recombination lines 
($\Delta$V$>$60~\kms), much broader than would be expected 
if the width were thermal ($\sim$25 \kms).  JMP refer to these
sources as Broad Line Recombination Objects (BRLOs). In their analysis,
JMP discuss the apparent correlation between spectral index values and 
radio recombination line (RRL) widths, summarize the existing models
that could explain the correlation, and conclude that the disk-wind
model (Hollenbach et al. 1994) is the most promising. Sewilo et al.
(2003) have recently carried out a VLA survey of seventeen hypercompact
(HC) HII regions, and report the detection of eight sources with H92$\alpha$
or H76$\alpha$ line widths greater than 40 \kms. They propose hierarchical
clumping in the nebular gas as an explanation of the spectral indices
observed.
Here we describe new H52$\alpha$ line
observations that have uncovered a number of BLROs in W49N consistent with
the 30\%-50\% of such sources found in these surveys.
The scales probed by these observations ($\sim$600 AU)
may further clarify the nature of BRLOs, and
provide new constraints for models of these young sources.

\section{OBSERVATIONS AND RESULTS} 
We have undertaken an extensive high frequency, 
high resolution imaging and spectroscopy program 
in the massive star forming region
W49A with the Very Large Array (VLA)
\footnote{The National Radio Astronomy Observatory is a facility of the National Science Foundation operated under a cooperative agreement by Associated Universities, Inc.} (Wilner et al. 2001, De Pree et al. 2000, De Pree, Goss \& Gaume 1998, De Pree, Mehringer \& Goss 1997). 
Observations presented here have been taken over the past 6 years, with the most recent observations made in early April 2001. 
For the high
resolution 7 mm images, we have combined continuum observations from 1995 (D configuration), 1998 (A configuration) and 2001
(B configuration). In the most recent B configuration observations, 7 mm receivers were present on 23 antennas, compared with only 12 in 1998
and 10 in 1995. The H52$\alpha$ line has a rest frequency
of 45.454 GHz, and all three
observations were made in the 1A mode with 25 MHz bandwidth,
centered at +8.0 \kms. Data were Hanning smoothed by the on-line system,
and spectra were separated into 15 channels, giving a channel separation and 
spectral resolution of
1.5625 MHz (10.3 \kms).

Observations were carried out with the fast switching calibration technique.
The
nearby bright calibrator source J1925+2106 was used for both phase and bandpass calibration (for the line data). 
The A, B, and D configuration data sets were independently calibrated, combined with the AIPS task DBCON and the final continuum image
was generated with the AIPS task IMAGR. Individual and combined continuum
images were self-calibrated. The resolution of the final 
continuum image is 0\farcs055 x 0\farcs044, PA=11.5$^o$, and the 
$rms$ noise is 0.34 mJy/beam. The $rms$ noise in the line channels
is 1.7 mJy/beam.
The corresponding resolution H52$\alpha$ spectral line data 
set was reduced using standard techniques in AIPS. 

Line parameters were generated from Gaussian fits to the 
data using the Groningen Image Processing System (GIPSY). 
Table 1 summarizes the results from the continuum and line data. 
Column (i) gives the source name, column (ii) 
indicates the peak line-to-continuum ratio, 
column (iii) gives the V$_{LSR}$
of the H52$\alpha$ line ($\theta_{beam}\sim$0\farcs05), 
column (iv) gives the H52$\alpha$ line width from the high resolution data, 
column (v) gives the H52$\alpha$ line width from the low resolution data 
(De Pree et al. 1997, $\theta_{beam}\sim$1\farcs7) and 
column (vi) gives the continuum spectral index.
Spectral index values in this table are obtained using the 1.3 cm and 3 mm 
flux densities, except as noted (Wilner et al. 2001). Errors in the spectral indices,
which are dominated by the absolute calibration of the continuum data,
are $\pm$0.1.

\subsection{Continuum Images} 
The upper left panel of Figure 1 shows the 3.6 cm continuum emission
from the central "ring" of HII regions in W49N (De Pree et al. 1997). Boxes
labelled (i) to (iv) are the locations of the four insets that show
the
full resolution ($\theta_{beam}$=0\farcs05) 7 mm continuum emission from the
ultracompact sources in W49N.
The addition of the B configuration data has improved the 7 mm image
quality over similar data presented by De Pree et al. (2000).
In particular, the additional $uv$ data provided by the 23 antennas in
the B configuration aided in the deconvolution of the beam carried out
by the AIPS task IMAGR.

We note that the interferometric 7 mm observations 
(even with the addition of the D
configuration data)
are missing flux density from emission on scales 
larger than about 40\arcsec.
The lower resolution imaging of W49A at 3.6 cm (De Pree
et al. 1997) clearly shows the presence of emission 
on scales that are not imaged in the 7 mm observations.
The comparison of these two images also emphasizes the
way in which high frequency observations are biased 
toward the detection of high-density gas.

\subsection{Line Data}
Figure 2 shows the H52$\alpha$ radio recombination line 
data and Gaussian fits from the 15 individual
sources in W49N from which we detect line emission. Vertical
bars in each spectrum indicate a 10\% line-to-continuum
ratio.
Parameters from fits to the
H52$\alpha$ line are given in Table 1 as described above. 
Note that the bandwidth available with the current VLA correlator at this high 
frequency is just sufficient to detect the broadest lines, and in some 
cases, the line widths quoted may be lower limits. Six of the 18
continuum sources have
H52$\alpha$ lines broader than 40 \kms, 
and four of these have lines broader than
50 \kms. 
Three of the sources have no detected line emission (C1, E and G2c).
In the case of C1 and E, the lack of  a line 
detection may be due to limited sensitivity,
since both sources are more extended, and source E
has line emission detected at 3.6 cm (De Pree et al. 1997). 
In the case of G2c, a high emission measure source with a rising spectral index,
the line may be too broad for the velocity coverage of the 7 mm observations with a
total velocity range of 155 \kms.

In Figure 2, sources G2a and G2b are shown with single Gaussian fits, 
and the parameters in Table 1 fit the line data with broad lines 
($\Delta$V$_{FWHM}>$50 \kms).
Two Gaussian fits (not shown) indicate that these 
sources may be double peaked profiles, with line peak separations of about 
33 \kms~and 38 \kms, respectively. We note that in the two-component fits,
the width of the individual peaks falls below 30 \kms.
The spectral resolution of these
observations is insufficient to be conclusive, 
but certainly these objects should be
reobserved when the new VLA correlator is available.

\section{DISCUSSION}
\subsection{Spectral Indices and RRLs}
The issue of impact broadening naturally
arises in high density sources like ultracompact HII regions.
Certainly, pressure (or impact) broadening is a 
significant effect in lower frequency ($n>$100) RRLs. 
However, the effect is strongly dependent on transition number ($n$), 
and less strongly dependent on the electron density. 

Using the equations presented in Brocklehurst \& Seaton (1972), the ratio of pressure broadening ($\Delta\nu^{I}$) to Doppler broadening ($\Delta\nu^{D}$) 
is 
\begin{equation}
{{\Delta\nu^I}\over{\Delta\nu^D}} = 0.142({{n}\over{100}})^{7.4}({{N_{e, true}
}\over{10^4}})
\end{equation}
where $n$ is the transistion number, and $N_{e, true}$ is the true electron density. Using the canonical value for the density of an HII region (10$^4$ cm$^{-3}$), it is clear that
pressure broadening is a small effect 
(${{\Delta\nu^I}\over{\Delta\nu^D}}<<1$) for transistions with n$<$100. 
Of course, the $N_{e, rms}$ density value that we use will be lower 
than the true value ($N_{e, true}$) in Eq. 1, dependent on the 
filling factor ($\epsilon$). If the filling factor is small, 
and the true electron density ($N_{e, true}$) much larger, 
then pressure broadening may become apparent at higher frequencies.

If the $N_{e, true}$ value becomes larger than 10$^4$ cm$^{-3}$, 
and we set the ratio of ${{\Delta\nu^I}\over{\Delta\nu^D}}$ in
Equation 1 equal to one, pressure broadening 
can become significant at smaller $n$ values (higher 
frequencies). For example, at densities of 10$^6$ cm$^{-3}$, 
pressure broadening
may become apparent at $n$=70.
At densities of 10$^{8}$  cm$^{-3}$, even the H38$\alpha$
line would experience pressure broadening. 

For the H52$\alpha$ line to be
pressure broadened would require true electron densities in the range of 10$^7$ cm$^{-3}$.  
The broad line H52$\alpha$ linewidths that are detected in W49N
are from regions which have $rms$ electron densities less than about
10$^{6}$ cm$^{-3}$, with most rms electron densities closer 
to 10$^{5}$ cm$^{-3}$ (De Pree et al. 2000). For these densities, 
pressure broadening would only be a significant effect for n$>$70.
And even if the filling factor ($\epsilon=({{N_{e~rms}}\over{N_{e~true}}})^2$) is as small as 0.01, pressure broadening would not be a 
factor at
45 GHz. However, if the filling factor is significantly smaller than 
0.01, then the broad lines detected may
be due to pressure broadening.
The high resolution of these observations also eliminates another 
possible explanation of broad lines: the overlap of sources with different $V_{LSR}$ values. Whatever the source of the broad lines, they arise from very
compact regions (d$\sim$600 AU).

It is possible to use the changing line width of the RRL lines as a function
of frequency to determine a lower limit for the $N_{e, true}$ for a particular
source.
For example, source D in W49N has a line width of 48$\pm$1.6 \kms~at 3.6 cm, 38.2$\pm$1.8 \kms~at 1.3 cm, and 35.1$\pm$1.2 \kms~at 7 mm (De Pree et al. 1997).
Using the 3.6 cm line width (48 \kms) for $\Delta\nu^I$ and the 7 mm line
width (35.1 \kms) for $\Delta\nu^D$, the lower limit for $N_{e, true}$ is
determined to be 1.8$\times$10$^5$ cm$^{-3}$, and the upper limit on the filling
factor is $\epsilon$=0.1. If the other sources in W49N have similar
filling factors, then pressure broadening is not likely to be the
major contributor to broad lines in these sources.

\subsection{Comments on Individual Sources}
There are 18 individual sources identified in the 7 mm continuum image (see Table 1). 
Of these, six sources (A1, A2, B1, B2, G2a \& G2b) have line widths 
greater than 50 \kms, and a seventh source 
with a rising spectral index (G2c)
has no line detected, perhaps due to its large line width.
Of the six sources with broad detected lines, four (B1, B2, G2a and G2b) have
rising spectra, with $\alpha >$0.8. Source A2 has a continuum spectral index of 0.3, and the fifth (A1) does not have a measured 
spectral index because only the 7 mm data has sufficient resolution
and sensitivity to detect the source.

The W49A core has between five and seven sources with characteristics that would classify them as BLROs, as defined by JMP. In their millimeter
survey, JMP found that approximately 30\% of UC HII regions were BLROs, and we 
find a consistent ratio of BLROs in the W49N core. 

\subsubsection{A Sources} Source A2 in W49N has `borderline' BLRO characteristics
(see Table 1).
As has been discussed in De Pree et al. (2000), the morphology of source A2  
is bipolar at lower resolution and frequency (3.6 cm image), and 
shell- or disk-like at higher resolution and frequency (7 mm image). 
Figure 1 shows the structures visible at different resolutions. The 
upper left panel shows the 3.6 cm data ($\theta_{beam}\sim$0\farcs8), and 
inset (i) shows the
7 mm continuum emission ($\theta_{beam}\sim$0\farcs055).  
Because the A configuration 7 mm data filters out large structures
($>$40\asec), the bipolar lobes are less apparent in the higher 
frequency data.  
Source  A is the most extended of the BLROs in W49N, and it may be near the end of its BLRO phase. 
However, it has a morphology and scale (see De Pree et al. 2000) 
that are consistent with the Hollenbach et al. (1994) photoevaporating disk 
model.  

\subsubsection{G Sources}
G1 and G1S are relatively extended, shell-like regions that have 
thermal line widths and continuum spectra consistent with an optically
thin HII region.
The G2a and G2b regions are located near the center of a high 
velocity molecular outflow traced in water maser emission (De Pree et al. 2000,
McGrath et al. 2003), and both have broad ($\Delta$V$>$ 50 \kms)
radio recombination 
lines 
and rising spectral indices. For comparison, the thermal broadening
of 10,000 K gas is 21.4 \kms. There is some indication that the 
lines in these two
regions are multiple-peaked (see Fig. 2), but higher spectral resolution observations are required. The doubly peaked lines may be related to the high velocity 
outflow. The G2c source is bright and elongated, but has no detected H52$\alpha$ line emission. The more extended G3, G4 and G5 sources 
(which seem to comprise portions of a large shell)
have line widths that are consistent
with thermal broadening, and spectral indices that are consistent with optically thin emission from ionized gas.

\subsection{UCHII Region Models}
JMP (1999) have reviewed the models currently proposed to 
explain the nature of UC HII regions. Originally these models were
required only to explain the morphology and longevity of UC HII regions. 
Observations in JMP and this investigation have shown that about 30\% of
UC HII regions have other characteristics that must be explained
by current models.
These properties include spectral index
values that are consistent with a constant velocity ($\alpha$=0.6) or 
accelerating ($\alpha>$0.6) ionized wind, and 
broad ($\Delta V_{FWHM}>$60 \kms) radio recombination
lines. 
In addition to the models reviewed in JMP, De Pree et al. (1995b) have
suggested that pressure confinement might explain the longevity of some UC HII
regions, and
Keto has described a gravitationally bound 
hypercompact HII region model that may also be relevant (Keto 2002a, 2002b).
Most recently,
Sewilo et al. (2003) have suggested that the rising spectral indices
detected in their HC HII region survey could be explained by hierarchical
clumping of nebular gas.

In their discussion, JMP conclude that while none of the models in their 
current form fully explain the new observations, the mass-loaded
wind (Dyson, Williams  \& Redman 1995) and the disk wind models
(Hollenbach et al. 1994)
are the most promising. As they point out, these two models can potentially 
account for the UC HII region lifetime problem. In addition, the
disk wind model can explain the r$^{-2}$ or steeper
density gradients that are suggested by the rising continuum
spectral indices, though other models (including champagne outflow)
could potentially account for this density profile.
Recent observational evidence for the presence
of disks around high mass stars may support the disk wind model. 
For example, 
near-IR observations (Conti \& Blum 2002) suggest that hot, dense disks might 
be present in the environments of very young massive stars in W49N. 
In addition, the clearly 
bipolar structure of source A2 in W49N observed at 
low frequencies (3.6 cm, see Fig. 1) is shown to have a
morphology consistent with an inclined 'disk' at high
frequencies (7 mm, see Fig. 1 inset (i)).
While sources with bipolar and disk morphologies are not 
widespread, the combined
ring/bipolar morphology in source W49N:A2 suggests that a
remnant accretion disk might be present in this source.

\section{CONCLUSIONS}
High resolution, high frequency line and continuum 
observations of the W49N region indicate that broad radio 
recombination line widths
detected on small scales at 1.3 cm persist at 7 mm. 
In addition, we have detected line emission from four sources (A1, B1, B3 \& G1S)
unresolved in previous H52$\alpha$
RRL observations of W49N (De Pree et al. 1995a). Six ($\pm$1) of the  
18 continuum sources in the W49N core have characteristics that would
classify them as BLROs, roughly in agreement with the 30\% value found in the
JMP survey of UC HII regions.
Broad lines in these regions detected at 7mm 
are likely to have a kinematic origin.
The new RRL observations clearly provide a discriminating factor 
between BRLOs and normal UC HIIs in this massive star forming region. 

\acknowledgements{CGD acknowledges the support of NSF grant AST-0206103, and
Jewels Deblasio for a careful reading of the manuscript.
DJW and CGD thank NRAO for
hospitality in Socorro, NM where some of this work was carried out. SK acknowledges the CONACyT grant 36568-E and DGAPA grant IN118401.}
  %
  %
  %
  %

  \clearpage 
  \begin{table} 
  \begin{center} 
  \begin{tabular}{lccccc}   \hline \hline
Source & ${T(L)}\over{T(C)}$ & V$_{LSR}$ &$\Delta$ $V_{HiRes}$ & $\Delta$ V$_{LowRes}$ & $\alpha$  \\ 
(W49N) &  & (km/s) & (km/s) & (km/s) & (S$_\nu$ $\propto$ $\nu^\alpha$)\\ \hline
A1  &  0.31$\pm$0.05 & 11.5$\pm$3.9 &  61$\pm$15 &. . .  & . . .\\ 
A2  &  0.28$\pm$0.03 & 15.4$\pm$1.8 &  43.4$\pm$5.4 & 47$\pm$2 &  0.3\\  \hline
B1  &  0.22$\pm$0.03 & 11.2$\pm$3.2 & 64$\pm$13  & . . . &  1.1\\ 
B2  &  0.21$\pm$0.02 & 16.8$\pm$1.9 & 43.9$\pm$5.8 & 49$\pm$3 & 0.9\\
B3  &  0.09$\pm$0.03 & 25.1$\pm$4.7 & 25$\pm$12 & . . . & . . . \\ \hline
C   &  0.40$\pm$0.02 & $-$8.0$\pm$0.7 & 32.4$\pm$1.8 &  39$\pm$3 & 0.1\\
C1  &  $<$0.50 &  &  &  & \\ \hline
D   &  0.73$\pm$0.07 & 6.5$\pm$1.3 & 28.6$\pm$3.4 & 35$\pm$1 & -0.1\\ \hline
E   &  $<$0.50 &  &  &  & \\ \hline
F   &  1.3$\pm$0.2 & 3.9$\pm$2.3 & 29.2$\pm$6.1 & 26$\pm$1 & -0.1\\ \hline
G1  &  0.57$\pm$0.05 & 2.9$\pm$1.1 & 29.2$\pm$2.9 & . . . & 0.3 \\
G1S &  0.94$\pm$0.1 & 14.2$\pm$1.7 & 24.1$\pm$4.5 & . . . & 0.0\tablenotemark{a}\\ 
G2a &  0.08$\pm$0.01 & 6.2$\pm$4.3 & 59$\pm$10 & 40$\pm$1 &  1.1\\
G2b &  0.13$\pm$0.03 & 22.8$\pm$5.7 & 58$\pm$14 & . . . & 0.8\\ 
G2c &  $<$0.05 &  &  &  & 0.9\tablenotemark{a}\\
G3  &  1.30$\pm$0.04 & 11.2$\pm$0.5 & 32.4$\pm$1.2 & . . . & 0.2\tablenotemark{b}\\
G4  &  1.60$\pm$0.09 & 9.4$\pm$0.8 & 27.7$\pm$2.0 & 34$\pm$1 & 0.2\tablenotemark{b}\\
G5  &  1.0$\pm$0.2 & 9.0$\pm$2.8 & 38.7$\pm$7.8 & . . . & 0.2\tablenotemark{b}\\ \hline
  \end{tabular} 
  \tablenotetext{a}{Spectral index between 7 mm and 3 mm} 
  \tablenotetext{b}{Spectral index between 3.6 cm and 3 mm}
  \end{center} 
  \tablenum{1} 
  \caption{H52$\alpha$ Line Properties and Spectral Indices}
\end{table}

\clearpage 
\begin{figure}[p] 
\vspace{18cm}
\caption{Upper left pseudocolor image shows the 3.6 cm radio continuum toward
W49N (from De Pree et al. 1997). Boxes numbered (i) through (v) indicate
the regions of detail showing the 7 mm radio continuum. The resolution
of the 3.6 cm data is $\sim$0\farcs8, and the resolution of the
7 mm data is 0\farcs055~x~0\farcs044 ($\sim$500 AU). Insets show
(i) W49N:A1 and A2, (ii) W49N:B1, B2 and D, (iii) W49N:C, (iv) W49N:G1, G2a, G2b and G2c, and (v) W49N: G3.
Offsets are from the phase center of the 7 mm
VLA observations ($\alpha$ 19 10 12.916, $\delta$ 09 06 11.916). 
The contour levels are at
-1, 1, 1.4, 2, 2.8, 4, 5.6, 8, 11.2, 16, 22.4, 32, and 44.8 times a 5$\sigma$
level of 1.7 mJy beam$^{-1}$.}
\end{figure} 

\clearpage 
\begin{figure}[p] 
\vspace{21cm}
\includegraphics{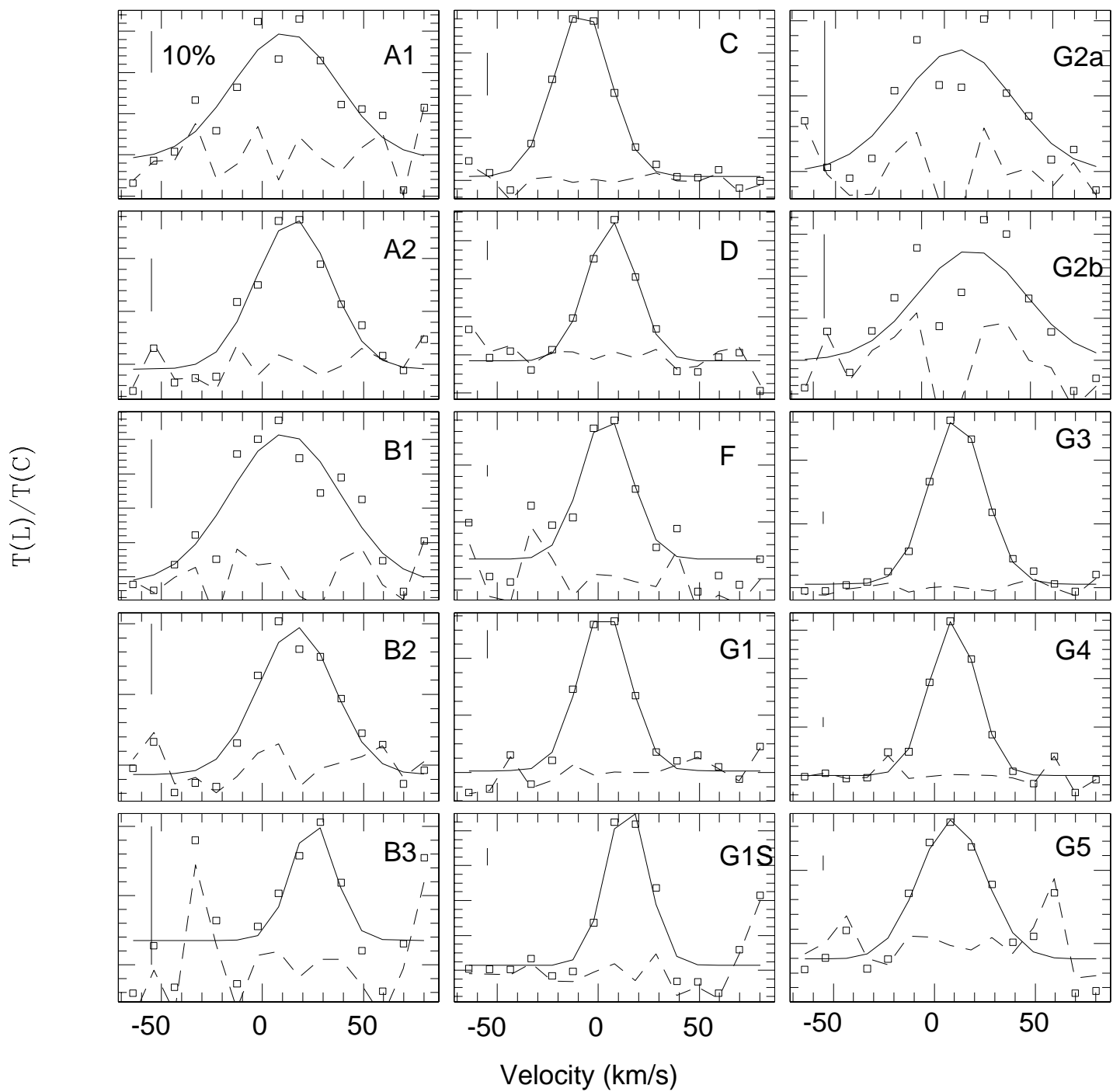} 
\caption{H52$\alpha$ (45.4 GHz) radio recombination lines from the individual 
sources. Plots show the data (open squares), single Gaussian
fit (solid line) and residual (dashed line). The parameters of the 
fits are given in Table 1. Vertical scale indicates line-to-continuum ratio,
and the vertical line in each plot indicates 10\%.}
\end{figure} 

\end{document}